# Electronically coupled complementary interfaces between perovskite band insulators

Mark Huijben[1], Guus Rijnders[1], Dave H.A. Blank[1], Sara Bals[2], Sandra Van Aert[2], Jo Verbeeck[2], Gustaaf Van Tendeloo[2], Alexander Brinkman[1], Hans Hilgenkamp[1]

[1]*Faculty of Science and Technology and MESA+ Institute for Nanotechnology, University of Twente, P.O. Box 217, 7500 AE Enschede, The Netherlands*

[2]*Electron Microscopy for Materials Research (EMAT), University of Antwerp, Groenenborgerlaan 171, 2020 Antwerp, Belgium*

**Perovskite oxides exhibit a plethora of exceptional electronic properties, providing the basis for novel concepts of oxide-electronic devices[1,2]. The interest in these materials is even extended by the remarkable characteristics of their interfaces. Studies on single epitaxial connections between the two wide-bandgap insulators LaAlO$_3$ and SrTiO$_3$ have revealed them to be either high-mobility electron conductors or insulating, depending on the atomic stacking sequences[3]. In the latter case they are conceivably positively charged. For device applications, as well as for basic understanding of the interface conduction mechanism, it is important to investigate the electronic coupling of closely-spaced complementary interfaces. Here we report the successful realization of such electronically coupled complementary interfaces in SrTiO$_3$ - LaAlO$_3$ thin film multilayer structures, in which the atomic stacking sequence at the interfaces was confirmed by quantitative transmission electron microscopy. We found a critical separation distance of 6 perovskite unit cell layers, corresponding to approximately 23 Å, below which a decrease of the interface conductivity and carrier density occurs. Interestingly, the high carrier mobilities characterizing the separate electron doped interfaces are found to be maintained in coupled structures down to sub-nanometer interface spacing.**

Perovskite oxides are commonly described in terms of their cubic unit cells, with the generic formula $ABO_3$. For a given parent compound, a rich phase diagram is colored-in by a substitution of the cations A or B, and/or a change in the oxygen stoichiometry. The ionic character of the chemical bonds and the consequent possibility of electronic reconstruction often render interfaces in these materials strongly electronically active. To understand this interface activity, it is instructive to describe the perovskites in terms of their constituting AO and $BO_2$ layering sequence. For example, whereas the two band-insulators $SrTiO_3$ and $LaAlO_3$ are seemingly similar, the SrO and $TiO_2$ layers are charge-neutral, while the charge states in the $LaAlO_3$ are $(LaO)^+$ and $(AlO_2)^-$, respectively. In heterostructures, the $AO-BO_2$ stacking sequence is maintained and consequently a polarity discontinuity arises at the $LaAlO_3-SrTiO_3$ interface. It has been shown by Ohtomo and Hwang[3] that due to this, the $LaO:TiO_2$ interface becomes conducting, and it is suggested that the conduction is governed by electron-transfer from $LaAlO_3$ into the $TiO_2$ bonds of the $SrTiO_3$. The complementary $AlO_2:SrO$ interface, with the $AlO_2$ possibly acting as an electron acceptor, remained insulating[3]. Such conducting interfaces are analogous to two-dimensional electron (hole) gases (2-DEGs) in semiconductors, which find applications in e.g., opto-electronic, high power RF and magnetoelectronic devices[4]. Great interest exists in the fundamental properties of electronically coupled 2-DEGs, placed very near to each other[5].

Until now the investigations on the $LaAlO_3-SrTiO_3$ interface conduction effects have concentrated on individual interfaces[3]. To study the electronic coupling of the complementary interfaces between these insulators, we fabricated high-quality multilayers in which a variable number of $LaAlO_3$ unit cell layers are stacked between $SrTiO_3$, and vice versa. The $LaAlO_3 -SrTiO_3$ heterostructures were grown by pulsed laser deposition, including in situ monitoring by Reflective High Energy Electron Diffraction (RHEED)[6]. Single-crystal $LaAlO_3$ and $SrTiO_3$ targets have been used, applying a KrF excimer laser at a repetition rate of 1 Hz and a laser fluency of ~1.3 J $cm^{-2}$. The deposition temperature was 850 $^o$C, and the oxygen pressure $3 \times 10^{-5}$ mbar.

The first type of heterostructure was deposited on $TiO_2$-terminated $SrTiO_3$ (100) substrates[7] and consisted of a $LaAlO_3$ layer followed by a $SrTiO_3$ top layer (Fig. 1a). Unit-cell RHEED intensity oscillations were used to control the number of unit cell layers for both materials. The thickness of the LaAlO$_3$ layer was varied from 1 to 26 unit cells while the number of unit cells for the SrTiO$_3$ top layer was always kept constant at 10. As the internal polarization of the LaAlO$_3$ is the driver for the interface doping, a change of the LaAlO$_3$ layer thickness could possibly result in a modification of this polarization. For this reason we have also fabricated heterostructures of a second type, in which a thin SrTiO$_3$ layer is sandwiched between sufficiently thick LaAlO$_3$ layers (Fig. 1b). For the fabrication of those heterostructures, the TiO$_2$-terminated SrTiO$_3$ (100) substrates were first covered with one monolayer of SrO by pulsed laser interval deposition[8] at 50 Hz. Subsequently, a LaAlO$_3$ base-layer of 13 unit cells was deposited, followed by a SrTiO$_3$ layer, of which the thickness was varied from 2 to 11 unit cells, and finished by a LaAlO$_3$ top layer of 13 unit cells. Atomic force microscopy of the completed heterostructures showed atomically smooth terraces separated by unit cell steps, similar to the original substrate surface.

To examine the atomic stacking sequence at the interfaces, a superlattice of LaAlO$_3$ and SrTiO$_3$ was grown on a TiO$_2$-terminated SrTiO$_3$ substrate. Figure 2a shows a high-angle annular dark field image (HAADF) taken along the [001] zone axis. In this image, which is obtained in a Titan 80-300 scanning transmission electron microscope (STEM), the intensities of the atomic columns scale with the atomic number Z. The weaker TiO and AlO columns are located in the center between respectively the brighter Sr and La columns. To investigate the atomic layers at the interface (e.g. AlO$_2$:SrO or LaO:TiO$_2$) the signal-to-noise ratio is increased by averaging over the area indicated by the white rectangle in Figure 2a. This part of the image is divided into smaller sub-sections, which are added after cross correlation in Figure 2b. To quantitatively evaluate the peak heights of the atom columns, statistical parameter estimation is used, in which a parametric model of Gaussian peaks is fitted to Figure 2b in the least-squares sense. Figures 2c and 2d show the estimated peak heights corresponding to the La, Sr, TiO, AlO columns and the atom columns ($X_A$, $X_B$) at both interfaces (indicated in Fig. 2b). Also the 90% confidence intervals have been computed using the so-called Cramér-Rao lower bound[9]. Comparing the confidence intervals corresponding to the peak height of the atom columns at the interfaces with those of the surrounding TiO/AlO peaks confirms that the interface on top of every SrTiO$_3$ layer is TiO$_2$:LaO, whereas at the bottom it is $AlO_2$:SrO. These configurations are also expected from the deposition parameters.

The electronic properties of the heterostructures were investigated by a four-point Van der Pauw method. For this, wire-bonded contacts were applied at the corners of the samples, connecting to the $(AlO_2)^-/(SrO)^0$ interface as well as to the $(LaO)^+/(TiO_2)^0$ interface. Measurements on individual interfaces using single $LaAlO_3$ layers on $SrTiO_3$ substrates confirmed the electron conduction of the $(LaO)^+/(TiO_2)^0$ interface with a sheet conductance of $1.4 \times 10^{-4}$ $(\Omega/\square)^{-1}$ at room temperature, while the $(AlO_2)^-/(SrO)^0$ interface had a sheet conductance of $\sim 10^{-7}$ $(\Omega/\square)^{-1}$. All single interface experiments showed photoconductivity due to photocarrier injection. Ultraviolet light illumination on single $LaAlO_3$ layers on $SrTiO_3$ substrates increased the conductivity by factors of 2 and 120 for the $(LaO)^+/(TiO_2)^0$ and $(AlO_2)^-/(SrO)^0$ interfaces, respectively. Conductance enhancements were only observed for wavelengths of the illuminated light below 380 nm, corresponding to the bandgap of 3.2 eV of $SrTiO_3$. To enable a careful analysis of the intrinsic interface coupling, the effects of photocarrier injection were suppressed in the multilayer studies by shielding the samples from any light during the experiments and the 24 hours before.

The sheet resistances $R_s$ at room temperature, for both types of heterostructures, are presented in Figure 3a for different values of the separation distance ($d$) between the two interfaces. A decrease in $d$ is found to be accompanied by an increase in $R_s$ below a separation distance of six unit cells, corresponding to 23 Å. Interestingly, both types of heterostructures show a similar dependence on the interface separation distance, albeit with a difference of 20% in the absolute value of $R_s$. The sheet carrier densities $n_s$ were deduced from measurements of the Hall-coefficient $R_H$, using $n_s = -1/R_H e$. The room temperature results are shown in Figure 3b. Below a separation distance of six unit cells a decrease in sheet carrier density occurs for both types of heterostructures. The constant $n_s$ for large $d$ has a value of $\sim 1.8 \times 10^{14}$ cm$^{-2}$, corresponding to a charge density of $\sim 29$ $\mu$C cm$^{-2}$, which is $\sim 0.27$ electrons per unit cell area on the $(LaO)^+/(TiO_2)^0$ interface. In this, the contribution by the $(AlO_2)^-/(SrO)^0$ interface to the carrier density is neglected, due to its much lower conductivity.

The change in $n_s$ and $R_s$ below a certain interface separation distance relates to the charge distribution in the heterostructure. There exists no theoretical modelling yet of the SrTiO$_3$ − LaAlO$_3$ interface, to our knowledge, but it is interesting to make the comparison with interfaces between the Mott insulator LaTiO$_3$ and the band insulator SrTiO$_3$.[10-12] Notably, for that system a characteristic distance of 6 unit cells was predicted, over which charge transfer and electronic reconstruction takes place[11,12]. The electronic reconstruction at a SrTiO$_3$-LaTiO$_3$ interface creates partially filled Ti-3d bands by band-bending effects and the symmetric confinement of charge leads to a suppression of $n_s$. An important difference with our experiments is the absence of TiO$_2$ layers in parts of our heterostructures. At the LaAlO$_3$ side, subbands are possibly created below the La-5d levels, in analogy with the Ti-3d levels at the SrTiO$_3$ side. The suppression of $n_s$, induced by the coupling between the interfaces, could then result from the constraint that the charge density at the $(AlO_2)^-/(SrO)^0$ interface is low.

The temperature dependence of the sheet resistance and the Hall coefficient provides further insight into the electronic properties of the interface electron gas. For the SrTiO$_3$/LaAlO$_3$/SrTiO$_3$ heterostructures the temperature dependence of the sheet resistance $R_s(T)$ is presented in Figure 4a for different values of the LaAlO$_3$ layer thickness ($d_{LAO}$) and Figure 4b shows the carrier density as determined from the Hall effect as function of temperature. The energy-scale over which charge carriers seem to be frozen out is 6.0 meV (Fig. 4c), which is comparable to observations in SrTiO$_3$ at low La-doping[13]. At temperatures above 100 K, the carrier density for $d_{LAO} = 2$ is approximately constant, for $d_{LAO} = 5$ the thermally activated increase continues.

The temperature dependence of the Hall mobility $\mu_H$ is given in Figure 4d. Above 50K the mobilities show a $T^{-2}$ power law dependence, characteristic of Fermi-liquid behavior. Electron-phonon interactions are typically weak in SrTiO$_3$ (as is known from the poor heat conduction) and would give rise to a Bloch-Grüneisen temperature dependence of the resistance, which is not observed. Although electron-electron scattering is typically suppressed by screening and the Pauli exclusion principle, it is known to be relevant in transition metals with partially filled d-shells. At the n-type interface, this effect is expected to be important, when interface electronic reconstruction makes the Fermi surface intersect the Ti-3d subbands[11,12,14]. For all heterostructures, $\mu_H$ at room temperature is found to be constant at $6.0 \pm 1.0$ cm$^2$ V$^{-1}$ s$^{-1}$, without any dependence on the separation distances ($d_{LAO}$ and $d_{STO}$). This value corresponds with room temperature mobilities reported for single interfaces[3,15]. The very large mobilities in the zero-temperature limit provide an estimated electronic mean free path of 100 nm – 1 $\mu$m. This large electronic mean free path and the fact that $\mu_H$ does not decrease for decreasing interlayer thickness indicates that electron scattering at impurities or crystalline defects in the nearby interface are not dominating effects for the mobilities at room temperature.

The studies on the LaAlO$_3$:SrTiO$_3$ heterostructures, presented above, prove the possibility to realize closely-spaced conducting sheets in these otherwise insulating oxide systems. This provides a perspective for novel all-oxide electronic devices as well as for basic studies. In this respect it is interesting to note that the employed growth techniques can also be applied to the fabrication of multilayers with exclusively n-type interfaces, by combining the SrTiO$_3$ and LaAlO$_3$ layers with single layers of LaTiO$_3$ or TiO$_2$.

### Acknowledgements

This work is part of the research program of the Foundation for Fundamental Research on Matter (FOM, financially supported by the Netherlands Organisation for Scientific Research (NWO)) and Philips Research. A.B., D.B., H.H. and G.R. acknowledge additional support from NWO. S.B. and S.V.A. are grateful to the Fund for Scientific Research-Flanders. The authors also acknowledge B. Freitag, A.J. Millis, Y. Ponomarev, F.J.G. Roesthuis, H. Rogalla, R. Wolters, W. van der Wiel, D. Veldhuis and FEI Company.

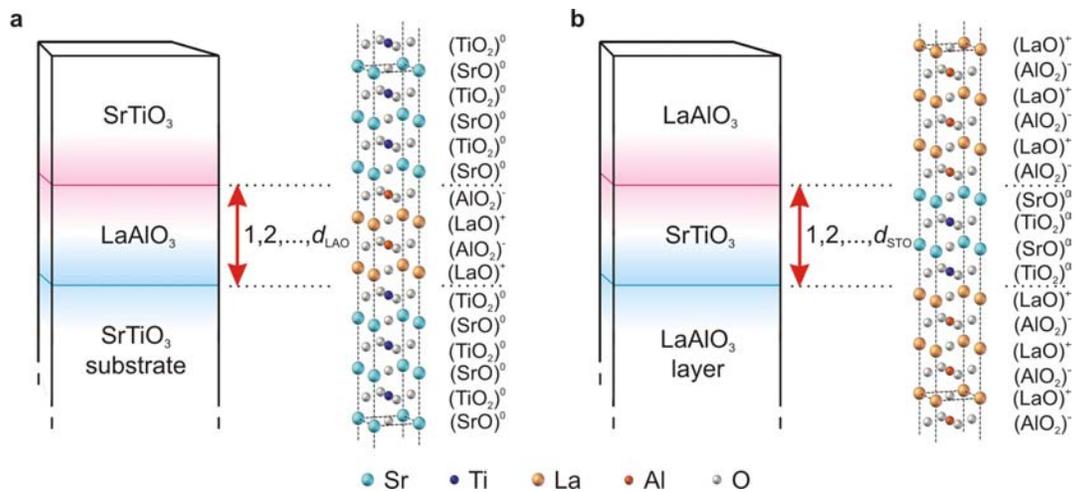

**Figure 1.** Representation of the investigated LaAlO$_3$/SrTiO$_3$ heterostructures. **a,** Schematic view of a SrTiO$_3$/LaAlO$_3$/SrTiO$_3$ heterostructure: a (001)-oriented LaAlO$_3$/SrTiO$_3$ bilayer grown on top of a TiO$_2$-terminated SrTiO$_3$ substrate, where the thickness of the LaAlO$_3$ layer ($d_{LAO}$) is varied. Atomic representation of the structure, showing the composition and charge state of each layer for the case of $d_{LAO} = 2$. **b,** Schematic view of a LaAlO$_3$/SrTiO$_3$/LaAlO$_3$ heterostructure: a (001)-oriented LaAlO$_3$/SrTiO$_3$/LaAlO$_3$ trilayer grown on top of a SrO-terminated SrTiO$_3$ substrate, where the thickness of the SrTiO$_3$ layer ($d_{STO}$) is varied. Atomic representation of the structure, showing the composition and charge state of each layer for the case of $d_{STO} = 2$.

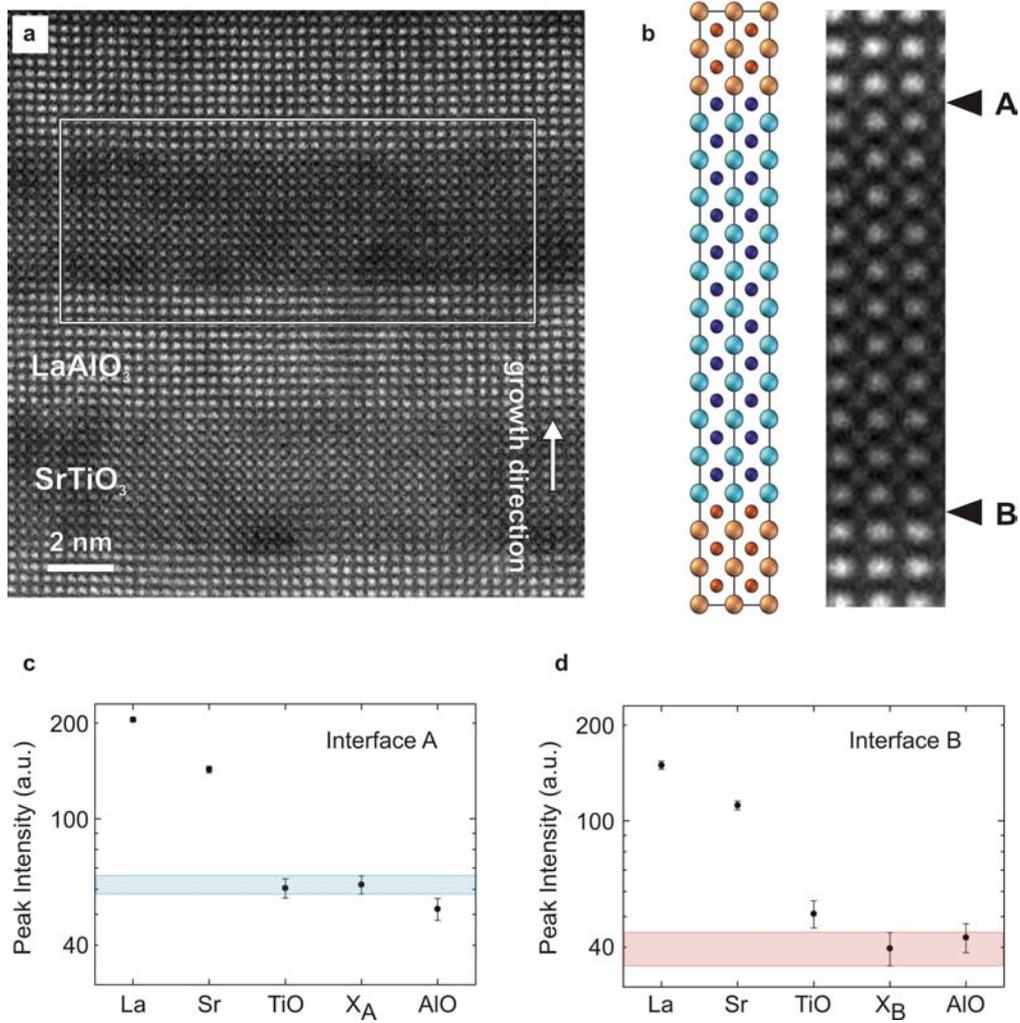

**Figure 2.** Quantitative scanning transmission electron microscopy analysis of the atomic stacking sequences at the interfaces. **a,** High-angle annular dark field (HAADF) image of the LaAlO₃/SrTiO₃ superlattice along the [001] zone axis. The area of the image indicated by the white rectangle is used for averaging. **b,** Image showing the result of the averaging procedure, together with the atomic representation of the structure. The atomic positions are indicated in size and color as in Fig. 1. The image is used as a starting point for quantitative peak estimation of the La, Sr, TiO, AlO columns and of the atom columns at the interfaces (indicated with A and B). **c+d,** The estimated peak heights of the La, Sr, TiO, AlO columns and of the atom columns at the interfaces (indicated with symbols $X_A$ and $X_B$) together with the corresponding 90% confidence intervals. The colored bands indicate the 90% confidence intervals of the atom columns at the interfaces ($X_A$ and $X_B$).

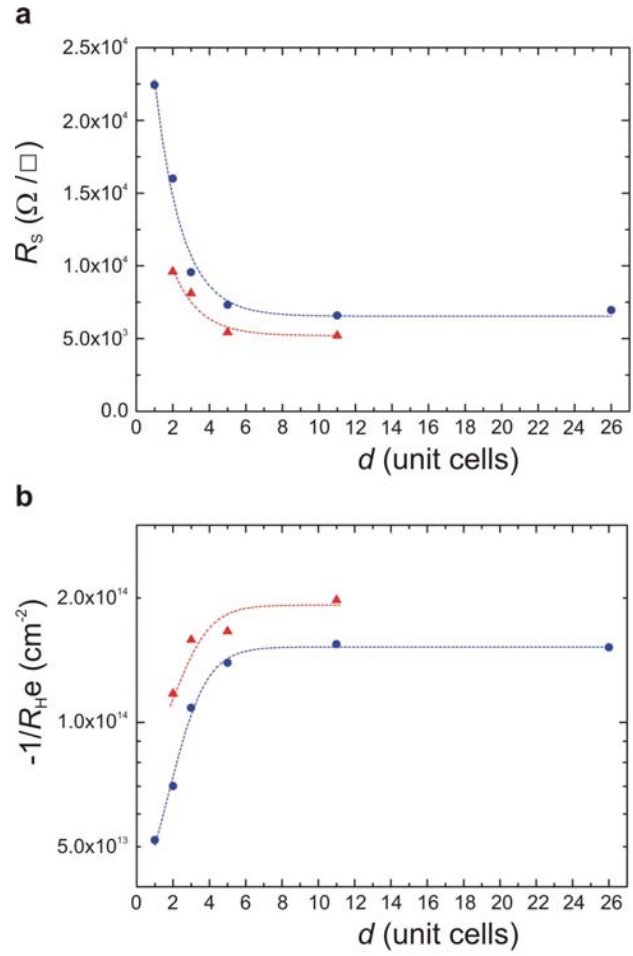

**Figure 3.** Electronic properties of the LaAlO₃/SrTiO₃ heterostructures at 300 K for different separation distances between the interfaces. **a,** Dependence of the sheet resistance $R_s$ on the separation distance $d$. **b,** Dependence of $-1/R_H e$ on the separation distance $d$. SrTiO₃/LaAlO₃/SrTiO₃ heterostructures and LaAlO₃/SrTiO₃/LaAlO₃ heterostructures are indicated by circles and triangles, respectively. The dashed lines are guides to the eye.

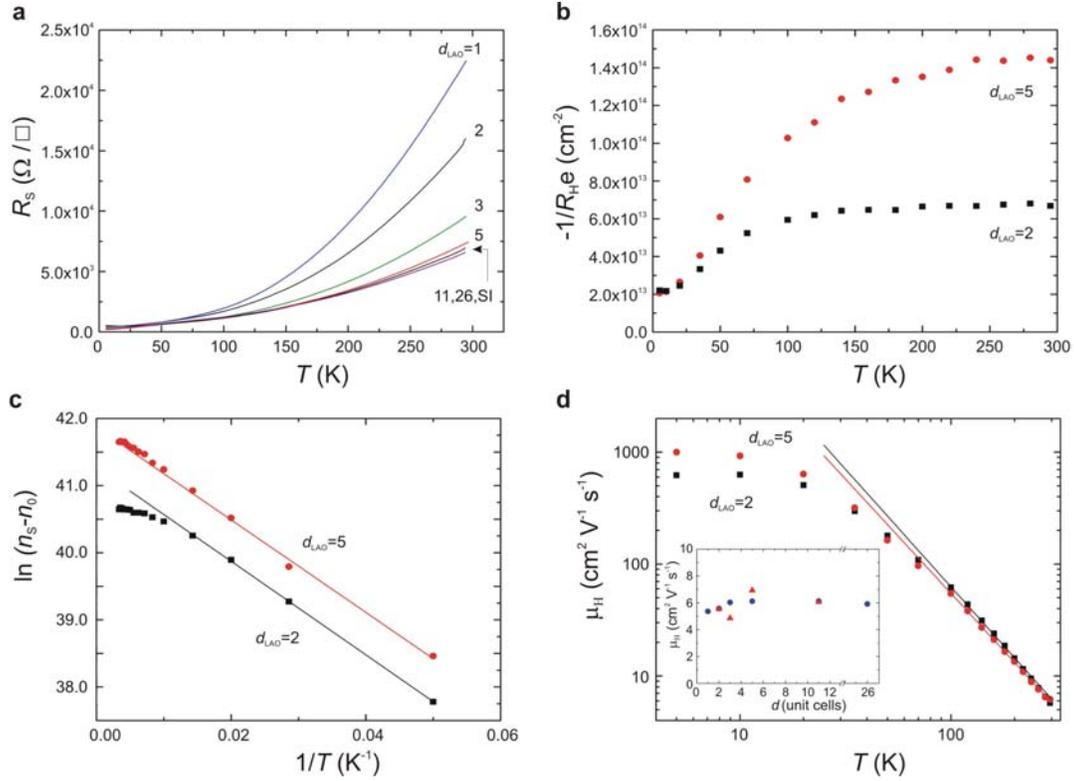

**Figure 4.** Transport properties of the LaAlO₃/SrTiO₃ heterostructures for different separation distances between the interfaces. **a,** The temperature dependence of the sheet resistance $R_s(T)$ for different thicknesses of the LaAlO₃ layer ($d_{LAO}$) in SrTiO₃/LaAlO₃/SrTiO₃ heterostructures. A measurement on a single $(LaO)^+/(TiO_2)^0$ interface (SI) is also indicated. **b,** Temperature dependence of $-1/R_H e(T)$ for SrTiO₃/LaAlO₃/SrTiO₃ heterostructures with separation distances of $d_{LAO}$ = 2 (squares) and 5 (circles). **c,** Temperature dependence of $\ln(n_S-n_0)$ for SrTiO₃/LaAlO₃/SrTiO₃ heterostructures with separation distances of $d_{LAO}$ = 2 (squares) and 5 (circles), where $n_S$ is defined as $-1/R_H e$ and $n_0$ is the low temperature limit of $n_S$. **d,** Temperature dependence of Hall mobility $\mu_H(T)$ for SrTiO₃/LaAlO₃/SrTiO₃ heterostructures with separation distances of $d_{LAO}$ = 2 (squares) and 5 (circles). The inset shows the separation distance $d$ dependence of $\mu_H(T)$ at 300 K, where SrTiO₃/LaAlO₃/SrTiO₃ heterostructures and LaAlO₃/SrTiO₃/LaAlO₃ heterostructures are indicated by circles and triangles, respectively.